\documentclass[12pt,preprint]{aastex}

\usepackage{emulateapj5}
\usepackage{onecolfloat}
\usepackage{graphicx}
\usepackage{times}

\def\gsim { \lower .75ex \hbox{$\sim$} \llap{\raise .27ex \hbox{$>$}} }
\def\lsim { \lower .75ex \hbox{$\sim$} \llap{\raise .27ex \hbox{$<$}} }

\newcommand{\Gaia}{{\it Gaia }}

\shorttitle{Streams in the nearby Galactic stellar halo}
\shortauthors{H. Koppelman et al.}

\begin{document}

\twocolumn[

\title{One large blob and many streams frosting the nearby stellar halo in \Gaia DR2}

\author{Helmer Koppelman\altaffilmark{1}, Amina Helmi and Jovan Veljanoski }
\affil{Kapteyn Astronomical Institute, University of Groningen,
P.O.Box 800, 9700 AV Groningen, The Netherlands.}

\begin{abstract}
  We explore the phase-space structure of nearby halo stars identified
  kinematically from \Gaia DR2 data. We focus on their distribution in
  velocity and in ``integrals of motion'' space as well as on their
  photometric properties. Our sample of stars selected to be moving at
  a relative velocity of at least 210 km/s with respect to the Local
  Standard of Rest, contains an important contribution from
  the low rotational velocity tail of the disk(s). The
  $V_R$-distribution of these stars depicts a small asymmetry similar
  to that seen for the faster rotating thin disk stars near the
  Sun. We also identify a prominent, slightly retrograde ``blob'',
  which traces the metal-poor halo main sequence reported by
  \citet{Babusiaux2018}. We also find many small clumps especially
  noticeable in the tails of the velocity distribution of the stars in
  our sample. Their HR diagrams disclose narrow sequences
  characteristic of simple stellar populations. This stream-frosting
  confirms predictions from cosmological simulations, namely that
  substructure is most apparent amongst the fastest moving stars,
  typically reflecting more recent accretion events.
\end{abstract}

   ]

\altaffiltext{1}{e-mail: koppelman@astro.rug.nl}


\section{Introduction}

The \Gaia 2nd data release \citep{Brown2018} has just become available
and has surpassed all expectations as evidenced by the science
verification publications accompanying its release
\citep[e.g.][]{Helmi2018,Katz2018,Babusiaux2018}. It will take many
years to fully exploit the vastness of the dataset and especially the
fantastic increase in accuracy. On the other hand, also because of the
same reasons, already a simple first exploration of the dataset yields
exciting new insights.

We report here the results of the analysis of the \Gaia DR2 set of 7
million stars with full phase-space information \citep[derived from
the astrometric and from the radial velocity spectrometer
data,][]{Lindegren2018,Katz2018rvs}, with the aim of identifying
substructure in the nearby Galactic halo. This Galactic component is
particularly important for understanding the assembly of the Milky Way
in a cosmological context.  It is here where we expect to find
merger debris and some of the most pristine stars
\citep{bj2005,2017MNRAS.471.2587S}. Large photometric surveys such as
SDSS and PanSTARRS have uncovered large overdensities and stellar
streams in the outer halo \citep{Belokurov2006,Bernard2016},
indicative of relatively recent accretion activity. However, the inner
regions of the stellar halo, despite containing most of the mass, have
remained more of a mystery so far partly because of the shorter
dynamical timescales. Yet it is these inner halo stars that tell us
about the early assembly history of the Milky Way.

In this {\it Letter} we analyze a sample of halo stars selected from a
Toomre diagram constructed from \Gaia DR2 data (Sec.~\ref{sec:data}). We have inspected their kinematics and
dynamics as well as their distribution in the Hertzsprung-Russell
diagram (Sec.~\ref{sec:results}). As reported in \citet{Babusiaux2018} two clear main sequences
are evident amongst halo stars in the Solar vicinity. Here we are able to associate at least in part the older and more metal-poor sequence to a prominent slighty retrograde
``blob", hinted at in previous datasets
\citep{Carollo2007,Morrison2009,Helmi2017} but never so easily discernible. We also
report the presence of several small clumps of stars that populate the
tails of the kinematic distribution, including the Helmi stream
\citep{H99}. Cosmological simulations of halo build-up
\citep{helmi2003} have long predicted that substructure due to accretion
should be more apparent at high-velocities \citep[see also][for first
hints]{2015AJ....150..128R}. Thus it is very plausible that these
substructures are in fact the remnants of the more recent accretion of
small dwarf galaxies contributing stars to the Solar neighborhood.

\begin{figure}
\centering
\includegraphics[width=8.5cm,clip=true]{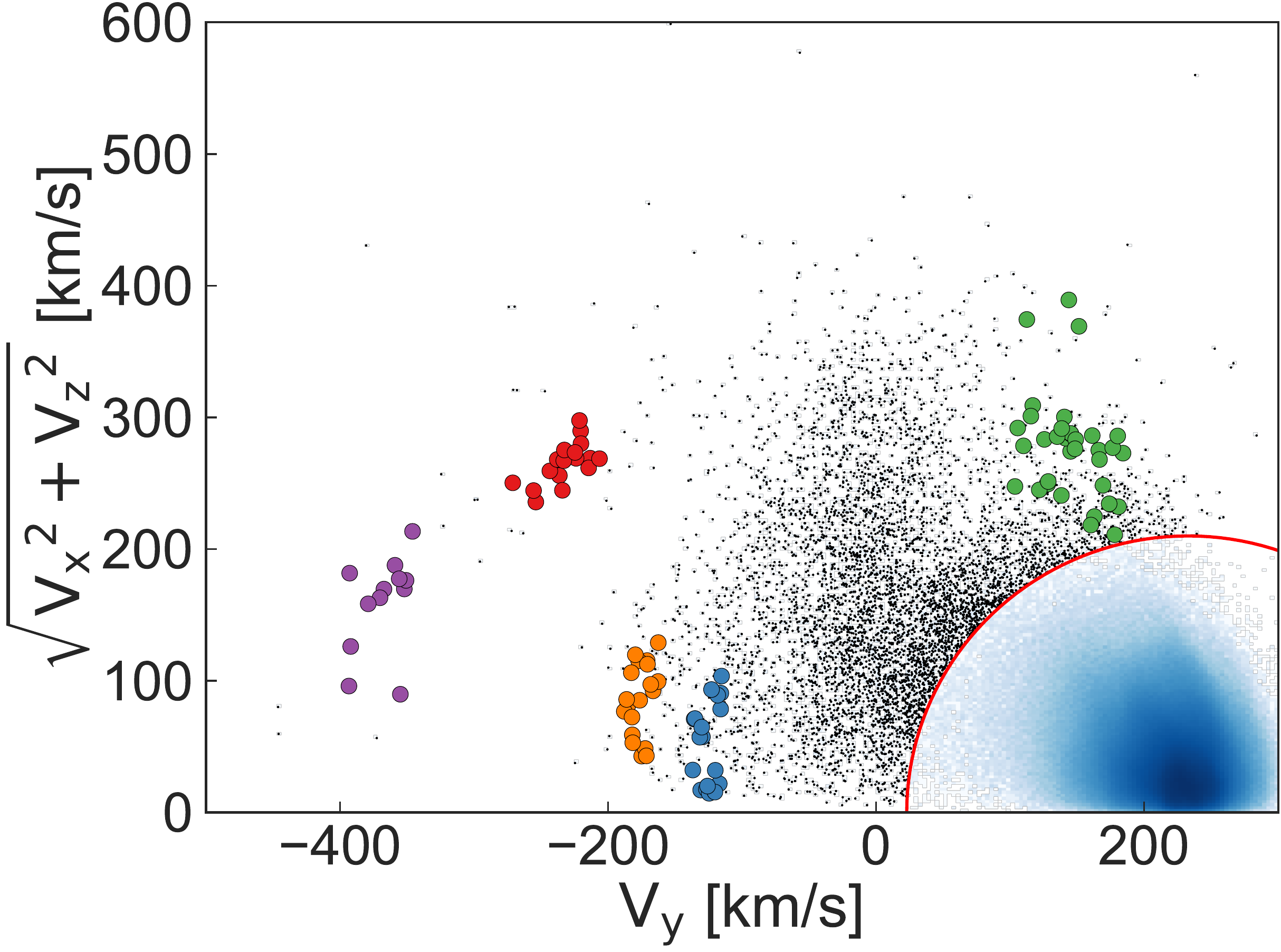}
\caption{Toomre diagram used to identify nearby halo stars (black dots and colored circles), defined as those
  that satisfy $|\mathbf{V} - \mathbf{V}_{\rm LSR}| > 210$~km/s, for ${V}_{\rm LSR} = 232$~km/s. The density map shows the distribution of all \Gaia DR2 stars within 1 kpc from the Sun. We consider only stars with relatively accurate parallaxes, i.e. with
  $\varpi/\epsilon_\varpi > 5$.}
\label{fig:toomre}
\end{figure}

\begin{figure*}[!ht]
\centering
\includegraphics[height=7cm,clip=true]{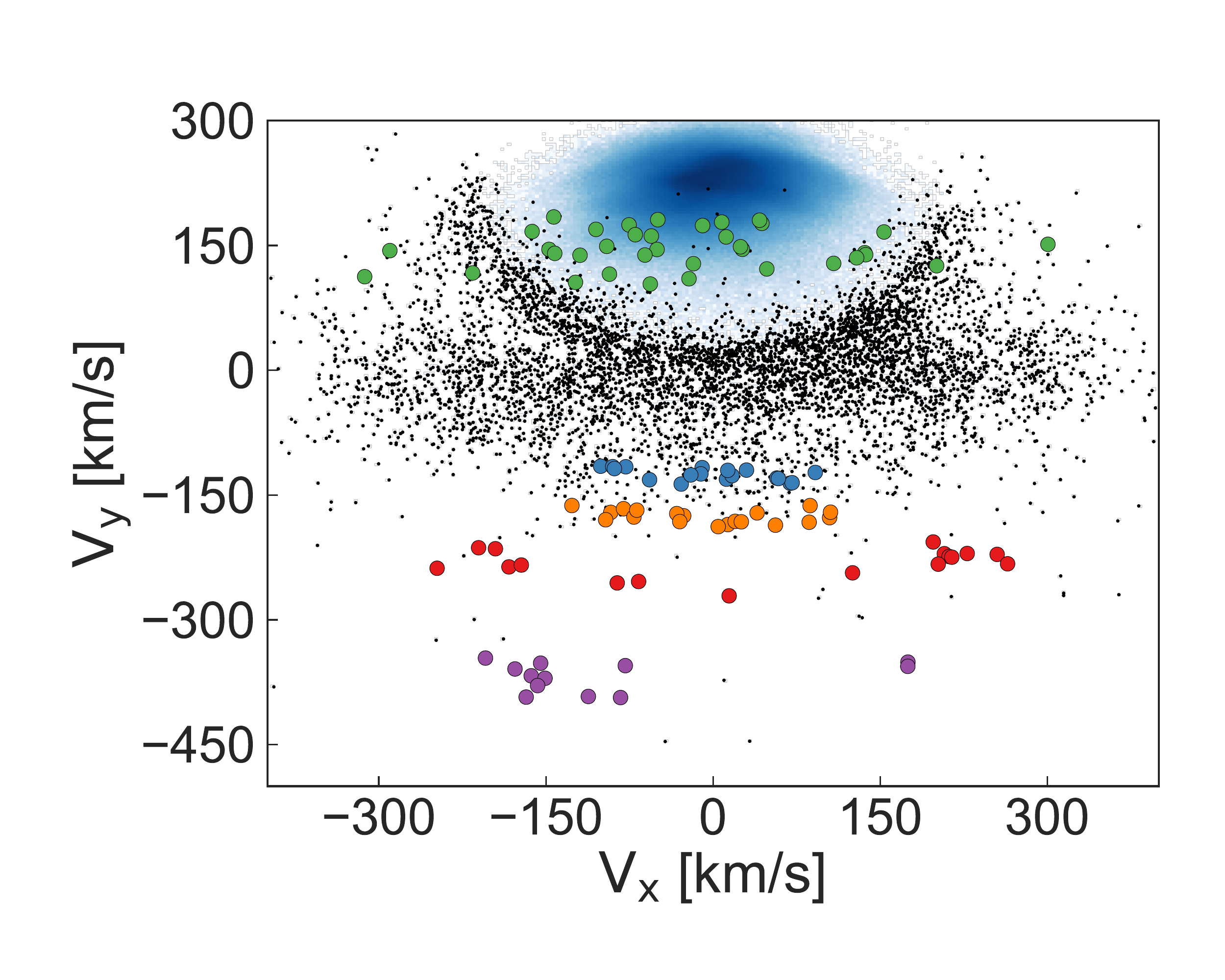}
\includegraphics[height=7cm,clip=true]{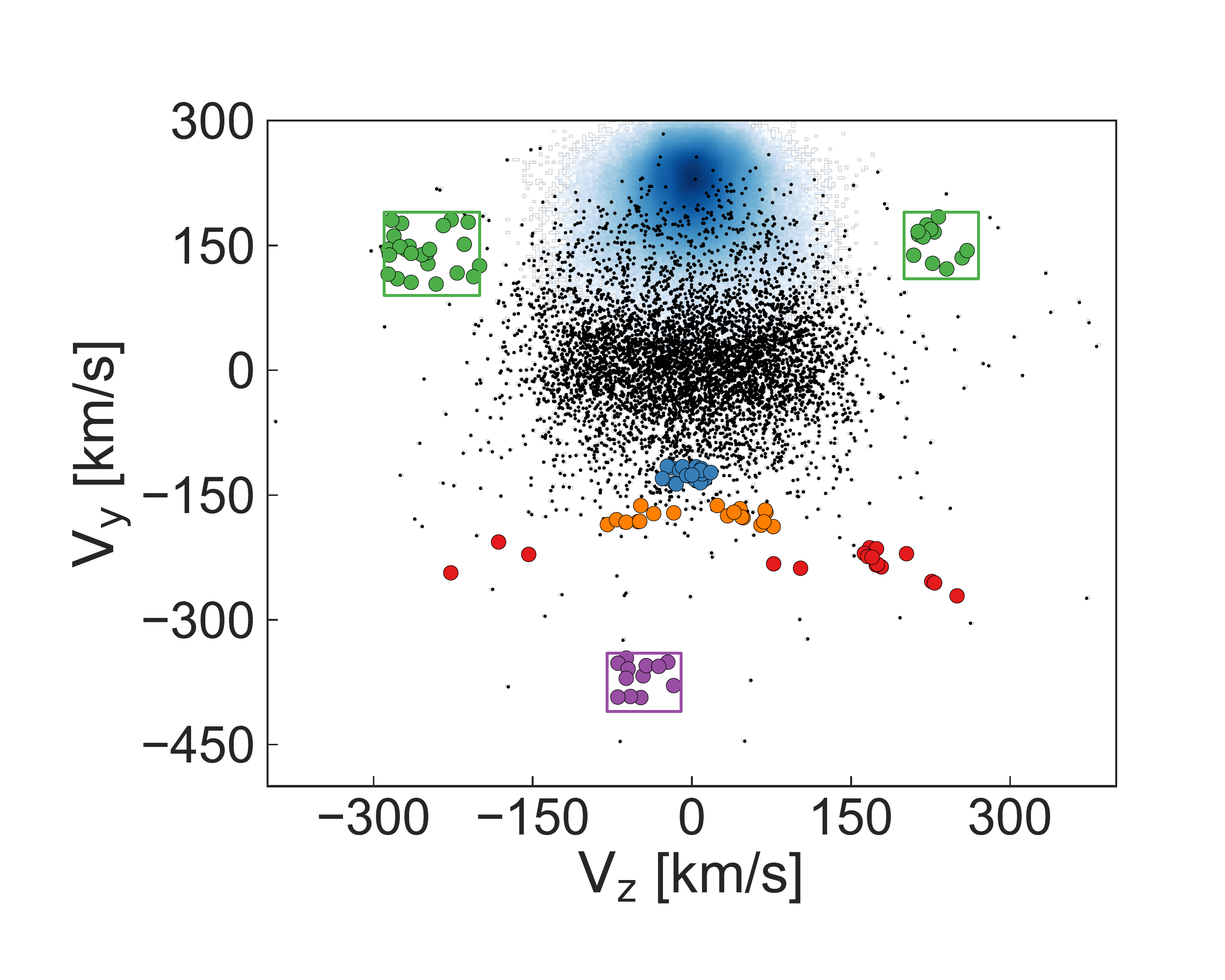}
\caption{Velocity distribution of halo stars (black dots and colored circles) selected according to the
  Toomre diagram shown in Fig.~\ref{fig:toomre}. The blue density maps show the
  velocity distributions of all stars within 1 kpc from the Sun and reveal the contribution of the disk(s).  The left panel exposes
  particularly clearly the effect of our kinematic selection criterion. The
  colored stars mark the location of tight clumps that are easy to discern because
  of their large velocities.}
\label{fig:vel}
\end{figure*}

\section{Data and Methods}
\label{sec:data}

Halo stars are relatively rare \citep[less than 1\% of the stars in the Solar neighborhood, e.g.][and references therein]{helmi2008}, hence it is paramount to have a good method of identification \citep[see][]{Velja2018}. Traditionally halo stars have been selected on the basis of their metallicity or their large velocities with respect to the disk \citep[see][for a recent comparison of selection methods]{Posti2017}. Here we follow the second approach, and use the traditional Toomre diagram selection to isolate halo stars \citep[see e.g.][]{2017ApJ...845..101B}. Although subject to biases (against halo stars that have similar motions as the disk(s)), it allows removing the large number of nearby disk stars that dominate the counts \citep{Brown2005}. 

The Toomre diagram plots the velocity in the direction of rotation
${V_y}$ against $\sqrt{{V_x}^2+{ V_z}^2}$. We use the \Gaia DR2 sample with 6D phase-space information and consider only those stars with relative parallax error
$\varpi/\sigma_\varpi > 5$, which allows us to compute distances as $d
= 1/\varpi$ with relative errors of 20\% at most. This sample contains 6366744 stars.

Fig.~\ref{fig:toomre} shows the Toomre diagram for the stars in this sample and located within 1~kpc from the Sun. Our reference system is oriented such that X is positive towards
Galactic longitude $l = 0$, Y in the direction of rotation, and Z for
Galactic latitudes $b > 0$, with the Sun located at $R_{\rm sun} = 8.2$~kpc on the negative X-axis. The velocities are also oriented in these
directions and have been corrected assuming a Local Standard of Rest
velocity $V_{\rm LSR} = 232$~km/s \citep{McMillan2017} and a peculiar motion for the Sun of $(U_\odot, V_\odot, W_\odot) = (11.1, 12.24, 7.25)$~km/s \citep{Schonrich2010}. 

We isolate a set of halo
stars using the criterion $|\mathbf{V} - \mathbf{V}_{\rm LSR}| > V_{\rm cut}$, where we take $V_{\rm cut} = 210$~km/s, \citep[i.e. slightly stricter than][]{Nissen2010}.
 This kinematically selected halo
sample contains a total of 5980 stars, with typical velocity
errors of 10 km/s. Of these, 3040 are main sequence stars. The colored solid circles in Fig.~\ref{fig:toomre} correspond to various easily discernable overdensities identified and discussed in more detail in the next section.

 \section{Analysis} 
\label{sec:results}

\subsection{Velocities}

Fig.~\ref{fig:vel} shows the velocity distribution of the
kinematically selected halo stars. This distribution is rather complex, particularly in the ${V_x} - {V_y}$ projection (left panel of Fig.~\ref{fig:vel}). This is in part due to the sharpness of the selection criterion applied on the Toomre diagram. This figure shows
also that relaxing slightly the value of  $V_{\rm cut}$ would lead to the inclusion of  more stars from the tail of the velocity distribution of the disk(s) \citep{SB2009,2017ApJ...845..101B}. Their imprint is a noticeable asymmetry in the \mbox{${V_x}$-distribution}, which is characteristic of the perturbation induced by the Galactic bar \citep[e.g.][]{Antoja2015}. In our sample, this asymmetry has a smaller amplitude than for the thin disk, but it is nonetheless clearly, and possibly unexpectedly, present at ${V_y} \gtrsim 100$~km/s, where the number of stars with ${V_x} < 0$ is 389, whereas for ${V_x} >  0$ there are 322 stars ($\sim 3\sigma$ excess). 
To properly understand the dynamical properties of this transition region of velocity space, would likely require a multi-dimensional probabilistic analysis to
assign stars to different physical components using also chemical and age information \citep{Binney2014, Posti2017}, which is beyond the scope of this {\it Letter}. 

The left panel of Fig.~\ref{fig:vel} shows a broad overdensity of stars for positive ${V_x}$ at ${ V_y \sim 50}$~km/s, just where the contribution from the low velocity tail of the disks would be expected to die away. Although to understand its nature requires an in-depth analysis, the  location of this overdensity seems unlikely to be related to the Toomre diagram-based selection criterion.

Fig.~\ref{fig:vel} shows the presence of a prominent component
with a large dispersion in ${ V_x}$ (equivalent to $V_R$ near the
Sun), that has a slightly retrograde mean motion of a few tens of km/s. 
Although this could be considered as the ``traditional" halo, it is slightly too retrograde and asymmetric towards more negative ${ V_y}$. Fig.~\ref{fig:vel} also reveals that the high-velocity tails of the 
distribution of the stars in our halo sample are populated by several cold
clumps. The structure at ${ V_y} \sim 150$ km/s and ${ V_z} \sim
-250$ km/s (in green, 25 stars), can be associated to one of the streams found
by \citet{H99}, and reported also in \cite{Helmi2018} using only
proper motion information (their Fig.~25). The second stream found by
\citet{H99} is less conspicuous with 12 stars, but present at ${ V_y} \sim 150$~km/s and ${ V_z} \sim 230$~km/s. The asymmetry in the number of stars in each of the streams implies that the accretion event from which these streams originate must have happened 6 to 9 Gyr ago according to the models presented in \citet{Kepley2007}. 

There are also other clumps in Fig.~\ref{fig:vel} and these are marked with different colors. Some appear to overlap with previously reported hints of
substructure \citep[e.g.][]{2015AJ....150..128R}. For example, the orange circles are probably related to the ``retrograde outlier stars" of \citet{Kepley2007}, and those with 
blue color overlap with the structure {\tt VelHel-4} from 
\citet{Helmi2017}. 

\begin{figure}
\centering
\includegraphics[width=8.5cm]{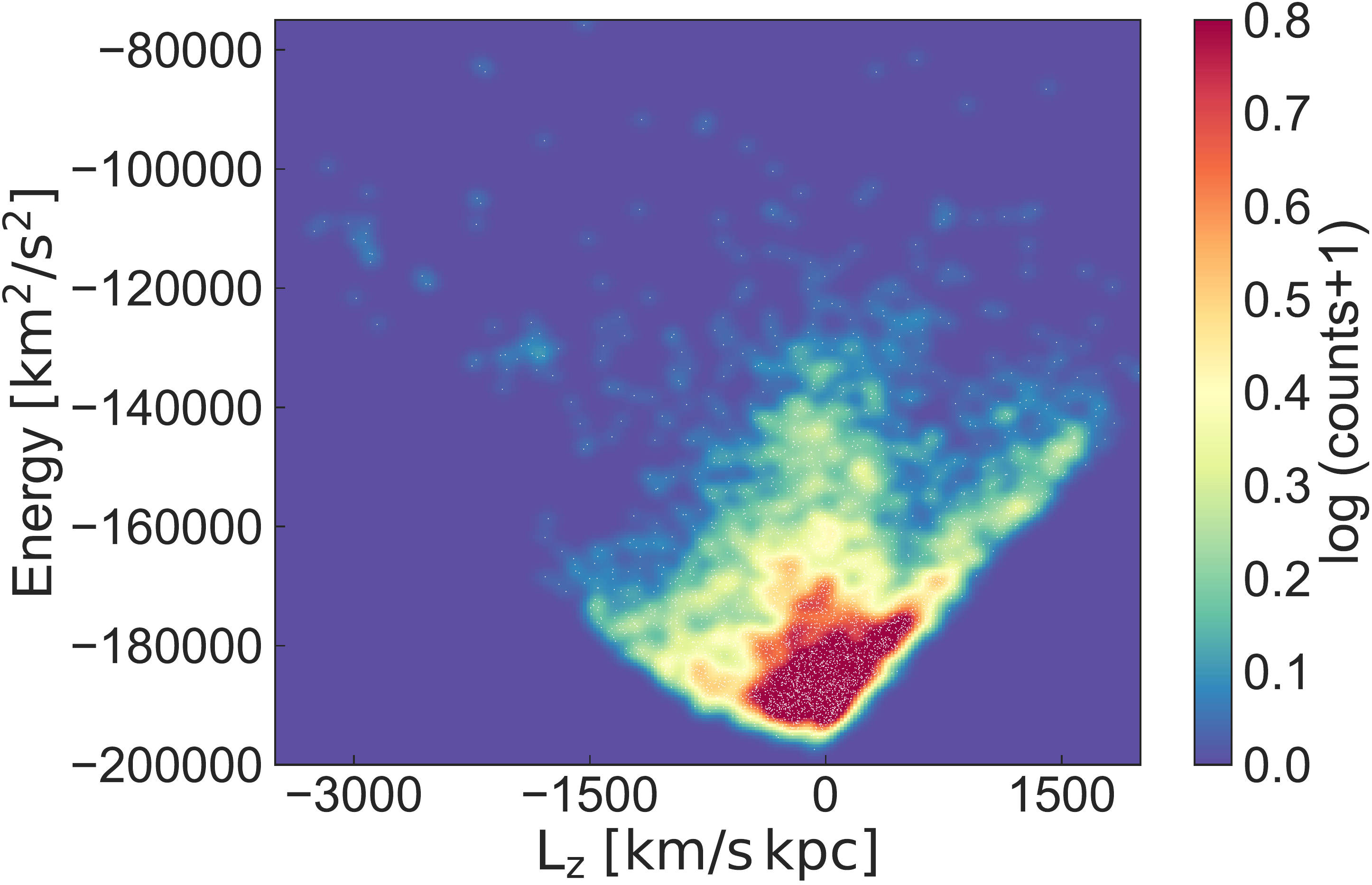}
\caption{Distribution of the stars in our kinematically selected halo sample in
  the ``integrals of motion'' space defined by their energy $E$ and
  $z$-component of their angular momentum ${ L_z}$. To make the structure visually more apparent we have used here a stricter value of ${ V_{cut} = V_{LSR}}$, which reduces the contrast between the tails of the disks and the rest of the halo.}
\label{fig:iom}
\end{figure}

\begin{figure}[!t]
\centering
\includegraphics[width=8.5cm]{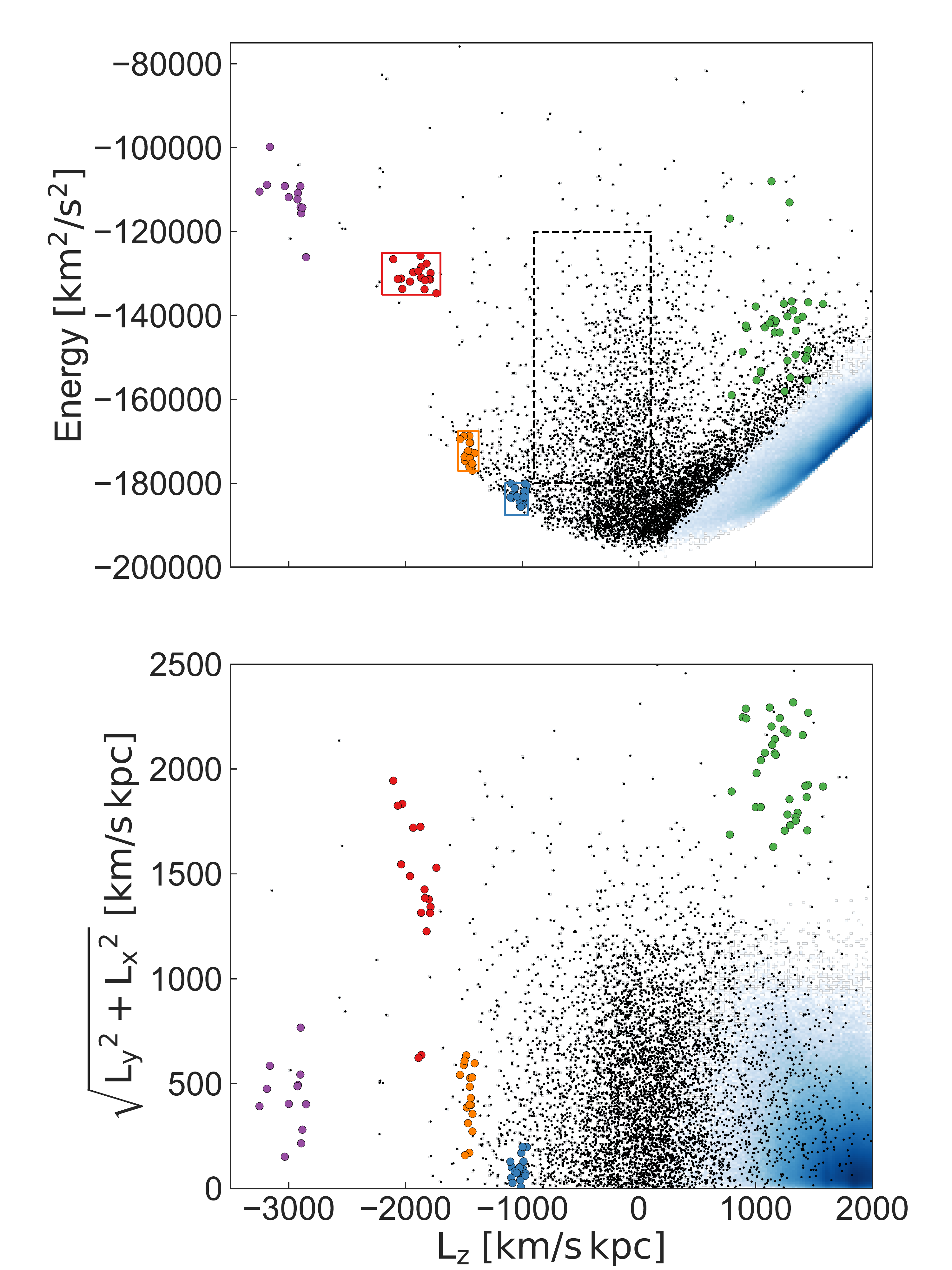}
\caption{Distribution of stars in energy $E$ vs ${ L_z}$ (top), and in ${ L_\perp}$ vs ${ L_z}$ (bottom), highlighting the location of the various tight structures also seen in Fig.~\ref{fig:vel}. The blue density maps mark the contribution of all the stars in the \Gaia DR2 6D sample located within 1 kpc from the Sun.}
\label{fig:iom-scatter}
\end{figure}

\subsection{``Integrals of motion''-space}

We explore now the distribution of stars in the space of ``integrals
of motion" defined by the $z$-component of the angular
momentum\footnote{For convenience we flip the sign of the $z$-angular
  momentum such that it is positive in the sense of rotation.} ${
  L_z}$, the perpendicular component ${ L_\perp =
  \sqrt{L_x^2+L_y^2}}$, and the energy $E$. For the stars in our
sample we compute their total energy $E$ as the sum of a kinetic and a
potential term, where the amplitude of the latter is estimated using a
suitable Galactic potential\footnote{The exact form and values of the
  characteristic parameters of this Galactic potential are not too
  relevant provided they yield a reasonable rotation curve for the
  Milky Way in the volume probed by the stars.} \citep[see][for
details]{Helmi2017}.  Note that $L_\perp$ is not really conserved in
an axisymmetric potential like that of our Galaxy, but it has proven
to be nonetheless a useful proxy for a third integral, and to help in
discriminating substructures with different orbital properties \citep{H99,2000MNRAS.319..657H}

Figure \ref{fig:iom} shows the distribution of the halo stars in our
sample in the $E - { L_z}$ space.  Although this figure is
reminiscent of that obtained using a TGAS$\times$RAVE sample
\citep{Helmi2017} or perhaps even TGAS$\times$SDSS
\citep{2018ApJ...856L..26M}, it is much more spectacular. 

Figure \ref{fig:iom} shows a
clear prominent ``blob'' or ``plume" that is slightly retrograde, the counterpart of the structure seen in velocity space (Fig.~\ref{fig:vel}) and also in the Toomre diagram (Fig.~\ref{fig:toomre}). This region has been previously associated to where a possible
progenitor of OmegaCen would deposit debris
\citep{2002ASPC..265..365D,2003MNRAS.346L..11B,Majewski2012,Helmi2017}. Whether
this is the only progenitor populating this region of phase-space
remains to be seen. Because of its large extent in energy, this structure contributes stars to the outer halo, and so may well be at least
partly responsible for the ``retrograde" component previously reported
by \citet[][and subsequent work]{Carollo2007}. If a single event, its size in ``integrals of motion" space suggests it was very
significant.

Fig.~\ref{fig:iom-scatter} shows a scatter plot of the distribution of kinematically selected stars in the $E - { L_z}$ (top) and ${ L_\perp - L_z}$ (bottom) spaces. We have plotted here
as a density map the contribution of all the stars in the \Gaia DR2 6D sample located within 1~kpc from the Sun and with distance errors smaller than 20\%. These figures clearly show where the disks fall and the sharpness of our kinematic selection criterion in the transition region between the disks and the (traditional) halo. 

Several tight overdensities also apparent in the panels of Fig.~\ref{fig:iom-scatter}. There is a close correspondence between these overdensities and those identified in
velocity space. The boxes shown here mark the clumps easy to identify in the ``integrals of motion'' space, while the remaining colored clumps have been identified in velocity space. 
All structures have been highlighted
 using the same color-coding as in
Fig.~\ref{fig:vel}. 

To establish the significance level of these structures we randomize the whole halo dataset by  reshuffling the velocities of the stars, while keeping their spatial distribution. We make 10.000 such random realizations and recompute for each, the distribution in ``integrals of motion" space. We find that in none of these realizations, a substructure is apparent that has a similar extent and location as any of the overdensities identified in the various figures.  

\subsection{HR diagram}

\begin{figure*}
\centering
\includegraphics[width=17cm]{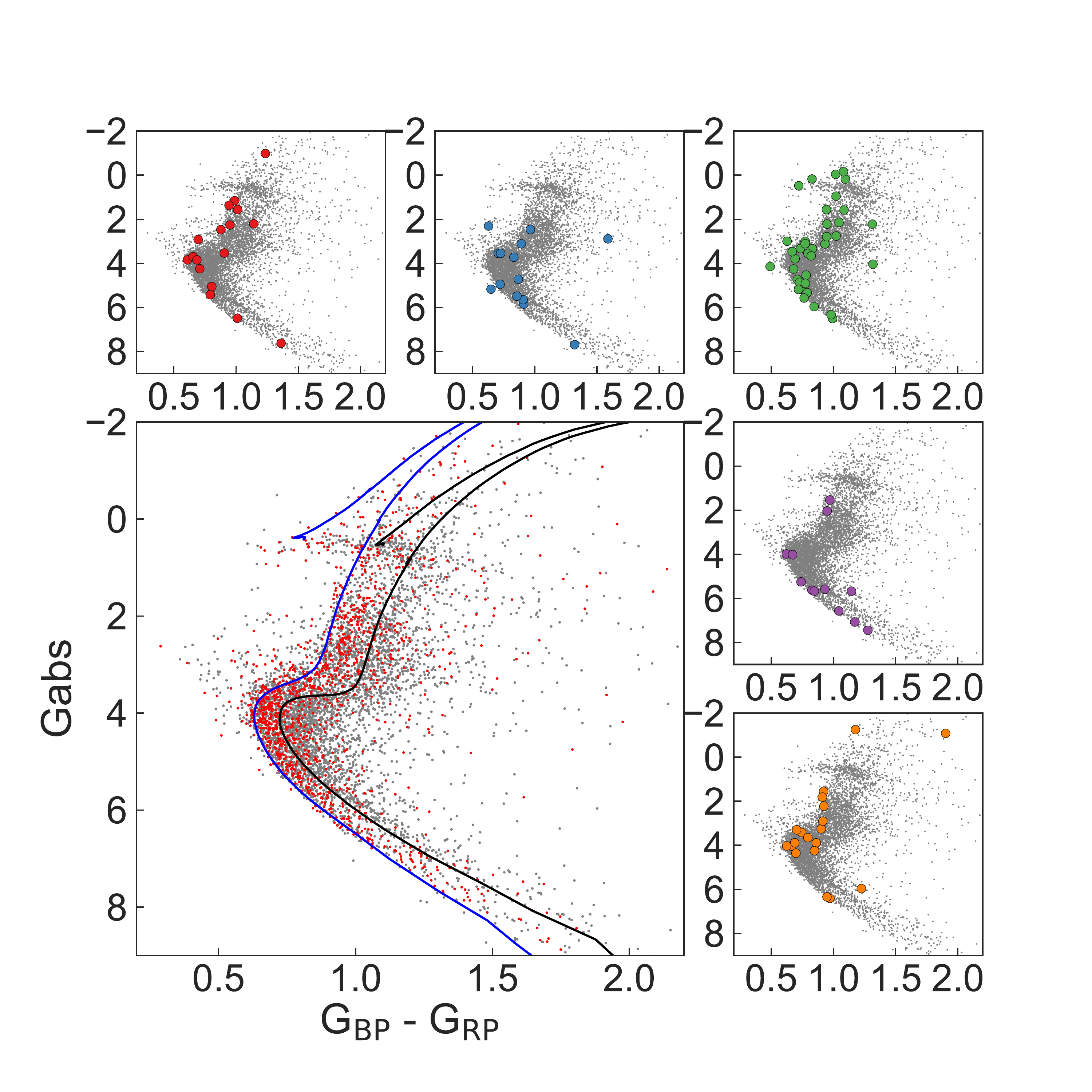}
\caption{HR diagrams of the stars in our kinematically selected halo sample (in gray). The location of the stars in the retrograde ``blob'' seen in Fig.~\ref{fig:iom} are shown in red in the large panel. The subpanels show the sequences defined by the other structures identified in Fig.~\ref{fig:vel}  and \ref{fig:iom-scatter} using the same color schemes. To guide the eye we include here two isochrones having [M/H] $\sim -1.3$ dex and 13 Gyr old (blue), and [M/H] $\sim -0.5$ dex and 11 Gyr old (black), following \citet{Babusiaux2018}.}
\label{fig:hrd}
\end{figure*}

The stars that are part of the large ``blob/plume'' identified in
Fig.~\ref{fig:iom} (inside the grey box in the top panel of
Fig.~\ref{fig:iom-scatter}) define a blue sequence in the HR diagram
of kinematically selected halo stars, as indicated by the red dots in
the large panel in Fig.~\ref{fig:hrd}. The absolute magnitude given here is calculated using the parallax from {\it Gaia} and has not been corrected for extinction. This sequence coincides with
that reported by \cite{Babusiaux2018}. These authors estimated an age of 13 Gyr and a metallicity [M/H] $\sim -1.3$ dex for this population, obtained by comparison to the blue isochrone shown here and obtained from \cite{Marigo2017}, considering a $\alpha$-enhancement of 0.23 \citep{Salaris1993}. The redder stars are fitted better by an isochrone with [M/H] $\sim -0.5$ dex and age of 11 Gyr (in black in Fig.~\ref{fig:hrd}), as already shown in \cite{Babusiaux2018}. 

The stars associated to the blue sequence (and the corresponding isochrone) are more metal-rich that the ``outer halo" component
of \cite{Carollo2007}, which has [Fe/H]~$\sim -2.2$~dex. This could imply that this region of
phase-space is more complex than anticipated \citep[see
also][]{Nissen2010}. On the other hand, if this overdensity is due to
the merger of a single large progenitor system, this could have had a
metallicity gradient. This would imply that stars at larger distances
could be more metal-poor on average \citep[as seen also for the
streams of the Sagittarius dwarf,][]{2006A&A...457L..21B}, thus
explaining the findings of \cite{Carollo2007}. 

The HR diagrams for the stars belonging to the other substructures are
plotted as small panels in Fig.~\ref{fig:hrd}. We do not attempt to fit isochrones to their distribution because of the relative low number of member stars (ranging from 12 to 37 stars). Note however, that their HR diagrams are very coherent, and suggest low metallicities and old ages, similar to those of the large ``plume" shown in the large panel of the figure. Exploration of the full \Gaia DR2 should disclose additional members. These could be either more distant giants stars or fainter nearby dwarfs. These dwarf stars would not have full phase-space information in \Gaia DR2 \citep[because of the magnitude cut of the spectroscopic sample, ][]{Katz2018rvs}, but should have accurate proper motions and parallaxes, and be present in large numbers.

\section{Discussion}
\label{section:disc}

\Gaia DR2 has revealed that the phase-space structure of halo stars
selected using a kinematic selection, namely in the Toomre diagram, is rather complex. It includes
large overdensities and several tight kinematic streams or clumps.

The distributions of stars in velocity and in ``integrals of motion"
space both indicate the presence of what appears to be the low
rotational velocity extension of the Galactic disks, well into the
region traditionally associated to the halo. Surprisingly, this
distribution is asymmetric in ${ V_x}$ ($V_R$) and follows the thin disk's
characteristic shape believed to be due to the effect of the Galactic
bar. This is presumably the metal-rich $\sim 11$~Gyr halo component reported in the HR diagram of \citet{Babusiaux2018}.  

We find a prominent slightly retrograde component, which we
interpret to be at least in part merger debris from one or more large
objects. The distribution of stars in the Toomre diagram shown in Fig.~\ref{fig:toomre} resembles very closely Fig.~7 of \citet{Villalobos2009} obtained from a simulation of the merger of a relatively massive object with a pre-existing disk (merger ratio 1:5), that subsequently gave rise to the formation of a  thick disk.  This large ``blob" could thus have been responsible for the puffing up of an ancient Galactic disk. An alternative
explanation is that this ``blob" is an ancient non-rotating halo,
and that the assumed Local Standard of Rest velocity should be shifted
by a few tens of km/s (from the 232 km/s we assume here), which at face value seems somewhat unlikely, especially given the asymmetric shape of the ``blob'' towards more negative rotational velocities. 

The velocity distribution of stars in our halo sample also reveals the presence of streams
located in the high-velocity tails, as predicted by cosmological simulations of the build-up of galactic halos \citep{helmi2003}. This implies that accretion
has played a role in the assembly of the halo near the Sun. How
important this process has been remains to be established with more
sophisticated analysis. Further explorations of \Gaia DR2 in other
regions of the Galaxy, using different tracers, and only proper motion
and parallax information are necessary to fully grasp the complexity
of the stellar halo. Given the superb quality of the data, there is no
doubt that \Gaia holds many surprises for those in the quest
to unravel the assembly history of the Galaxy.

\begin{acknowledgements}
We gratefully acknowledge financial support from a VICI grant to AH from the Netherlands Organisation for Scientific Research, NWO. This work has made use of data from the European Space Agency (ESA)
mission \Gaia (\url{http://www.cosmos. esa.int/gaia}), processed by the 
\Gaia Data Processing and Analysis Consortium (DPAC,
\url{http://www.cosmos.esa.int/web/gaia/ dpac/consortium}). Funding for the DPAC
has been provided by national institutions, in particular the institutions
participating in the \emph{Gaia} Multilateral Agreement. We thank Maarten Breddels for his data visualization and exploration tool: {\tt vaex} \citep{Breddels2018}.
\end{acknowledgements}

%
%

\bibliographystyle{apj} 

\end{document}